# Pressure/temperature/substitution-induced melting of A-site charge disproportionation in Bi$_{1-x}$La$_x$NiO$_3$ (0 ≤ x ≤ 0.5)

S. Ishiwata,[1,*] M. Azuma,[1,2] M. Hanawa,[2,3] Y. Moritomo,[2,3] Y. Ohishi,[4] K. Kato,[4,5] M. Takata,[4,5] E. Nishibori,[3] M. Sakata,[3] I. Terasaki,[6] and M. Takano[1]

[1]Institute for Chemical Research, Kyoto Univ., Gokashou, Uji 611-0011, Japan

[2]PRESTO, Japan Science and Technology Agency (JST), Kawaguchi, Saitama 332-0012, Japan

[3]Department of Applied Physics, Nagoya Univ., Furo-cho, Chigusa-ku, Nagoya 464-8601, Japan

[4]JASRI, Japan Synchrotron Radiation Research Institute, 1-1-1 Kouto, Mikazuki-cho, Sayo-gun, Hyogo 679-5198, Japan

[5]CREST, Japan Science and Technology Agency (JST), Kawaguchi, Saitama 332-0012, Japan

[6]Department of Applied Physics, Waseda Univ., Ookubo, Shinjuku, Tokyo 169-8555, Japan

**Metal-insulator transitions strongly coupled with lattice were found in Bi$_{1-x}$La$_x$NiO$_3$. Synchrotron X-ray powder diffraction revealed that pressure ($P \sim 3$ GPa, $T = 300$ K), temperature ($T \sim 340$ K, $x = 0.05$), and La-substitution ($x \sim 0.075$, $T = 300$ K) caused the similar structural change from a triclinic (insulating) to an orthorhombic (metallic) symmetry, suggesting melting of the A-site charge disproportionation. Comparing crystal structure and physical properties with the other $A$NiO$_3$ series, an electronic state of the metallic phase can be described as [$A^{3+}\underline{L}^{\delta}$, Ni$^{2+}\underline{L}^{1-\delta}$], where a ligand-hole $\underline{L}$ contributes to a conductivity. We depicted a schematic $P$-$T$ phase diagram of BiNiO$_3$ including a critical point (3 GPa, 300 K) and an inhomogeneous region, which implies universality of ligand-hole dynamics in $A$NiO$_3$ ($A$ = Bi, Pr, Nd…).**

## I. INTRODUCTION

Charge ordering is a common phenomenon in mixed-valence transition-metal oxides, but it attracts much attention because of competition with fascinating metallic behavior exhibiting superconductivity or giant magnetoresistance.[1,2,3,4] The competition between them has been keenly discussed as a clue for understanding the origin of such striking properties.[2,3] In addition to the mixed-valence systems, several integer-valence perovskite oxides such as $CaFeO_3$ and $ANiO_3$ ($A$ = Y, Pr, Nd...) also show a charge ordering transition described as $2M^{n+} \rightarrow M^{(n-\delta)+} + M^{(n+\delta)+}$, which is called charge disproportionation (CD).[5,6,7,8,9] CD can be detected as a symmetry breaking from orthorhombic ($GdFeO_3$-type structure) to monoclinic symmetry, which is caused by the breathing-type cooperative displacements like $O\text{--}M^{n+}\text{--}O\text{--}M^{n+}\text{--}O \rightarrow O\text{---}M^{(n-\delta)+}\text{---}O\text{-}M^{(n+\delta)+}\text{-}O$.

The MI transition in perovskite $ANiO_3$ series, associated with CD, has been studied systematically by substitution of A-site and/or application of pressure.[10,11,12,13,14] The MI transition temperature, $T_{MI}$, decreases monotonically as the size of an A-site ion or external pressure increases, i.e., the conduction band made of $3d$(Ni) and $2p$(O) orbitals becomes wider as the mean Ni-O-Ni bond angle increases. We should note here that the electronic properties of oxides containing an unusually high-valent ion such as $Fe^{4+}$ and $Ni^{3+}$ may be dominated by an oxygen-hole character which stems from strong $p$-$d$ hybridization.[15,16,17] Photoemission spectroscopic studies suggest that the realistic electronic state of $Ni^{3+}$ in $(NiO_6)^{9-}$ octahedron is close to $Ni^{2+}\underline{L}$ ($\underline{L}$ : a ligand-hole; a hole in a ligand oxygen $2p$ orbital).[16] CD in $AMO_3$ could thus be expressed as $2M^{n+}\underline{L} \rightarrow M^{n+} + M^{n+}\underline{L}^{2\delta}$, where ligand-holes play a crucial role.

Recently we have found a novel fashion of CD in a triclinic (space group; $P$-1) perovskite $BiNiO_3$. The crystal structure analysis indicates that the valence state is not $Bi^{3+}Ni^{3+}O_3$ nor $Bi^{3+}Ni^{(3-\delta)+}_{0.5}Ni^{(3+\delta)+}_{0.5}O_3$ but $Bi^{3+}_{0.5}Bi^{5+}_{0.5}Ni^{2+}O_3$.[18] The divalent nature of Ni ions and the triclinically distorted structure, caused by the A-site CD, lead to insulating behavior. Since "$Bi^{5+}$" have a very deep open 6s level, and such a high valence state at A-site is not stable from the viewpoint of Madelung potential energy[19], the realistic charge configuration of $Bi^{5+}$ should be expressed as $Bi^{3+}\underline{L}^2$. That is, ligand-holes are trapped and ordered in the Bi-O sublattice rather than in the Ni-O sublattice. $Bi^{4+}$ is a common ion showing CD as reported in $BaBiO_3$, where an antibonding 6s(Bi) - 2p(O) conduction band produces superconductivity by an appropriate substitution.[20,21] In the case of $BiNiO_3$, a competition between $Bi^{3+}\underline{L}$ ($Bi^{4+}$) and $Ni^{2+}\underline{L}$ ($Ni^{3+}$) is expected to exhibit unprecedented localization-delocalization transitions of ligand-holes. Indeed, we have observed a structural transition of $BiNiO_3$ to an orthorhombic symmetry at 513 K.[18,22] However, the physical properties of the orthorhombic phase is unclear because of a partial decomposition due to a thermal oxygen loss.

In this paper, we report structural and physical properties of $Bi_{1-x}La_xNiO_3$ as functions of pressure ($x = 0$), temperature ($x = 0.05$), and La - content ($0 \leq x \leq 0.5$). These data consistently demonstrate melting of A-site CD, accompanied with insulator (triclinic) to metal (orthorhombic) transitions dominated by ligand-hole dynamics.

## II. EXPERIMENT

Polycrystalline samples of $Bi_{1-x}La_xNiO_3$ ($x = 0, 0.05, 0.075, 0.1, 0.2, 0.5$) were obtained by high pressure syntheses as described before.[18] A precursor was prepared by dissolving stoichiometric amounts of $Bi_2O_3$, $La_2O_3$, and Ni in nitric acid, followed by heating at 730 °C in air for 6 h. A mixture of the precursor and an oxidizer $KClO_4$ (20 wt. % to the presursor) was treated at 1000 °C and 6 GPa for 30 min. The obtained samples were washed in distilled water to dissolve KCl. The samples used for measurements of resistivity were pressed to be dense at 6 GPa at room temperature.

Powder XRD data for phase identification were recorded on a Rigaku RINT 2500 diffractometer using CuKα radiation. SXRD data of $BiNiO_3$ under high pressure and these of $Bi_{1-x}La_xNiO_3$ at ambient pressure (AP) were collected using a diamond anvil cell (DAC) at beam line BL10XU and a large Debye-Scherrer camera at BL02B2 (Ref.[23]) of SPring-8, respectively. The granularity of the powder samples were homogenized to 2-3 μm in diameter by the precipitation method. For DAC experiments ethanol/methanol mixture was used as a pressure transmitting medium. To reduce the absorption effect of Bi ions, we selected as short wavelength as $\lambda = 0.4966$ Å for DAC experiments and $\lambda = 0.42084$ Å for structure analyses of La-substituted samples. The diffraction data for $Bi_{0.95}La_{0.05}NiO_3$ were taken between 100 K and 400 K with $N_2$ gas flow apparatus at 0.776 Å. All the data were analyzed by the Rietveld method using a Rietan 2000 program.[24]

Pressure dependence of DC resistivity was measured by a two-probe method using a cubic-anvil-type high-pressure apparatus. Electrical resistivities between 2 K and 400 K were measured by a four-probe method using a Quantum Design PPMS at a rate of 2 K/min. Thermopower was measured by a steady-state technique with typical temperature gradient of 1 K/mm, and the contribution of the voltage leads was carefully subtracted. DC magnetic susceptibility measurements were performed with a Quantum Design MPMS XL SQUID magnetometer in an external magnetic field of 0.1 T on cooling.

## III. RESULTS

First of all, we survey several XRD patterns indicating structural phase transitions caused by pressure, temperature, and La-substitution (Fig. 1(a)-(c)). In Fig. 1(a), 5 main peaks in the diffraction pattern at the bottom, characteristic of the triclinic cell, decreased to 3 of the orthorhombic cell; a structural transition from AP-phase (ambient pressure) to HP-phase (high pressure). The XRD pattern of HP-phase at 3 GPa was indexed assuming an orthorhombic unit cell of $5.32 \times 5.50 \times 7.62$ Å ($GdFeO_3$-type cell). The similar behavior can be seen in the SXRD patterns for $Bi_{0.95}La_{0.05}NiO_3$ taken at various temperatures (Fig.1(b)). LT-phase (low temperature) and HT-phase (high temperature) were indexed with triclinic (*P*-1) and orthorhombic (*Pbnm*) symmetries, respectively (see Fig. 2). The $GdFeO_3$-type orthorhombic structure ensures melting of CD in the HP-phase and HT-phase, for there is only one equivalent Bi site. Fig. 1(c) shows the XRD patterns of $Bi_{1-x}La_xNiO_3$ taken at room temperature. With increasing La-content, $x$, the number of the main peaks changes gradually from 5 to 3 via the composition with $x = 0.075$, where these two phases coexist.

Figure 3(a)-(d) shows evolutions of structural parameters under high pressure with traversing phase transition from triclinic (insulating) to orthorhombic (metallic) symmetry. The unit cell volume $V$ and the lattice parameters $a$, $b$, and $c$ decrease steeply across the transition to HP-phase (volume change at 3 GPa is 2.5 %). Since the unit cell volume of perovskite type structure is dominated by the *B*-O bond distance, the mean Ni-O distance of the HP-phase is expected to be shorter than that of the AP-phase. On increasing pressure, triclinic angles, $\alpha$, $\beta$, and $\gamma$, tend to merge into 90°, but still not unified at 3GPa. In accordance with the structural phase transition, pressure-dependent resistivity of $BiNiO_3$ at room temperature ($T = 300$ K) shows a steep drop by several orders of magnitude around 3 GPa, suggesting that an external pressure causes delocalization of ligand-holes trapped in the Bi-O sublattice (Fig. 3(d)).[25]

The temperature-dependent variations of unit cell volume, lattice parameter, unit cell angles, and resistivity of $Bi_{0.95}La_{0.05}NiO_3$ are quite similar to those as function of pressure (Fig. 4(a)-(d)). The discontinuity and coexisting of two phases on the verge of 340 K indicates the first-order nature of this transition. The magnitude of the volume contraction at 340 K ($\Delta V/ V = -3$ %) is comparable to that between the AP-phase and the HP-phase of $BiNiO_3$. Focused on the structural and the transport behavior, both increasing temperature and pressure induce melting of CD and delocalization of carriers in the same manner.

By controlling La-content, $x$, in $Bi_{1-x}La_xNiO_3$, similar plots of structural parameters and resistivity can be seen at 300 K (Fig. 5(a)-(d)). On increasing $x$, the resistivity decreases by 5~6 orders of magnitude with the decrease of unit cell volume, $\Delta V/V = -3$ %, as reported previously (Fig. 4(d)).[26] La-substitution for $x = 0.075$ leads to symmetrical change from the triclinic to the orthorhombic phase, which is metallic without CD, corresponding roughly to the applied pressure of 3 GPa. However, the SXRD data indicates the presence of a small amount of the triclinic phase for $x = 0.1$ (16.7 % wt. fraction), whereas no triclinic phase can be detected for $x = 0.2$.

As can be seen in Fig. 6(a), the La-substituted samples show metallic behavior, and diffuse MI transitions with large thermal hystereses are observed for $x \leq 0.1$. The samples with $x = 0.075$ and 0.1 show the reentrant metallic behavior at low temperatures, suggesting incompleteness of the MI transition. The thermopower, $S$, for $x = 0.075$ and 0.1 are roughly proportional to $T$ at low temperatures as observed in the metallic compound $LaNiO_3$ (Fig. 6(b)).[27] This behavior is by contrast with that for $x = 0.05$ which diverges below $T_{MI}$, as a hallmark of an insulator. We will discuss this point later.

Figure 6(c) and 6(d) show the temperature dependence of susceptibility and inverse susceptibility of $Bi_{1-x}La_xNiO_3$, measured in a magnetic field of 0.1 T on cooling. Surprisingly, all the compositions undergo antiferromagnetic ordering near 300 K, and the Neel ordering temperature, $T_N$, seems to have no correlation with $x$, which is a remarkable dissimilarity to the other members of $ANiO_3$. Below $T_N$, ferromagnetic moment due to canted spins emerges. In addition, the paramagnetic susceptibility for all samples seemingly obeys the Curie-Weiss law as shown in Fig. 6(d).

So far, we have demonstrated that increasing pressure, temperature, and La-substitution lead to the orthorhombic phase where A-site CD is absent. Here, we take a look at the detailed structural features to compare them quantitatively with each other. Figure 7(a) summarizes GdFeO$_3$-type distortion of $A$NiO$_3$, represented by $b/a$ where $a$ and $b$ denote lattice parameters (BiNiO$_3$ has also GdFeO$_3$-type unit cell with triclinic distortion). The value $b/a$ decreases linearly as the A-site ion becomes larger from Y to Pr (taken from Ref. 10), i.e, as the tolerance factor, $t$ (= $(r_A + r_O) / \sqrt{2}(r_{Ni} + r_O)$), becomes larger. Suppose the valence state of the A-site ion is trivalent, a linear relationship is expected between $b/a$ and $r_A$ as plotted in Fig. 7(a). However, Bi$_{1-x}$La$_x$NiO$_3$ ($x$ = 0.1, 0.2) and BiNiO$_3$ show the remarkable upward deviation from the expected line (even larger than SmNiO$_3$), which indicates that the effective tolerance factors are smaller than the expected values assuming the trivalent ionic radii for each cations. This deviation is derived from the misestimation of its charge distribution; i.e., the realistic electronic state for Bi$_{1-x}$La$_x$NiO$_3$ should be denoted as $[A^{3+}\underline{L}^{\delta}, Ni^{2+}\underline{L}^{1-\delta}]$ ($A$ = Bi$_{1-x}$La$_x$). $\delta$ is equal to 1 for $x$ = 0 (BiNiO$_3$). $\delta$ decreases with increasing $x$, and eventually $\delta$ is supposed to be zero when $x$ become 1 (LaNiO$_3$). Remaining of ligand holes in A-site is actually confirmed by the mean Ni-O bond distances of Bi$_{1-x}$La$_x$NiO$_3$, 1.973 Å ($x$ = 0.2) ~ 2.091 Å ($x$ = 0); see Table I), which are significantly longer than the predicted value for Ni$^{3+}$-O$^{2-}$ (1.937 Å). By using the variation of the indicator, $b/a$, we tried to make quantitative comparison between La-content, $x$, and pressure, $P$ (Fig. 7(b)). The plots clearly show equivalency of these parameters, allowing us to interchange La-content with chemical pressure.[28]

## VI. DISCUSSIONS

Our scenario deduced from structure analysis, resistivity, and susceptibility measurements is as follows. The AP-phase of BiNiO$_3$ and the LT-phase of Bi$_{0.95}$La$_{0.05}$NiO$_3$ have the same structure and the electronic state, described as $[A^{3+}_{0.5} + A^{3+}\underline{L}^2_{0.5}, Ni^{2+}]$ ($A$ = Bi or Bi$_{0.95}$La$_{0.05}$); an antiferromagnetic insulator due to the divalent nature of Ni ion. Increasing pressure, La-content, or temperature suppresses the development of CD as confirmed by symmetry change to orthorhombic perovskite, giving rise to conductive behavior. The steep drop of the resistivity under high pressure has clearly indicated that $BiNiO_3$ is rather conductive when CD is absent. Although we failed to determine the atomic fractional coordinates for HP-phase, a lattice contraction accompanied with the symmetry change across the phase transition suggests that the electronic configuration of HP-phase is the same as orthorhombic phase of $Bi_{1-x}La_xNiO_3$. As proposed in Fig. 7(a), we describe the electronic state of the orthorhombic phase as $[A^{3+}\underline{L}^{\delta}, Ni^{2+}\underline{L}^{1-\delta}]$ ($A = Bi_{1-x}La_x$).

Next, let us compare structural and physical properties of the orthorhombic phase of $Bi_{1-x}La_xNiO_3$ with that of $SmNiO_3$ ($T_{MI}$ = 400 K, $T_N$ = 225 K). Structural analyses based on SXRD were performed for the samples with $x$ = 0 (given in Ref. 18), 0.05 (see Fig. 2), 0.1, and 0.2, the refined structure parameters being listed in Table I. In comparison with $SmNiO_3$, the mean Ni-O-Ni angles of $Bi_{1-x}La_xNiO_3$ (151.4 ~ 153.4°, 0.05 ≤ $x$ ≤ 0.2) are comparable or even smaller ($SmNiO_3$; 153.2°), and the mean Ni-O bond distance (1.973 ~ 1.986 Å) is longer ($SmNiO_3$; 1.952 Å)[29], which suggests that the conduction band made of 3$d$ and 2$p$ orbital is smaller than that of $SmNiO_3$. Nevertheless, $Bi_{0.8}La_{0.2}NiO_3$ keeps metallic-like behavior far below $T_{MI}$ of $SmNiO_3$. This is ascribable to a strong hybridization between 6$s$(Bi) and 2$p$(O), being consistent with the presumed electronic configuration, $[A^{3+}\underline{L}^{\delta}, Ni^{2+}\underline{L}^{1-\delta}]$ ($A = Bi_{1-x}La_x$). Consequently, the insulator to metal transition in $Bi_{1-x}La_xNiO_3$ should be expressed as $[A^{3+}_{0.5} + A^{3+}\underline{L}^2_{0.5}, Ni^{2+}] \rightarrow [A^{3+}\underline{L}^{\delta}, Ni^{2+}\underline{L}^{1-\delta}]$. As suggested by the Ni-O bond length, considerable amount of ligand-holes exist around the A-site even in the orthorhombic phase, which bears localized spins of $Ni^{2+}$ ($S$ = 1). This appears to be consistent with the composition-independent $T_N$ of $Bi_{1-x}La_xNiO_3$. Of course, the possibility of the mixing of magnetic (insulating) triclinic phase in the metallic orthorhombic phase as the reason for the composition independent $T_N$ cannot be ruled out at this stage. Such mixings were indeed detected by the XRD studies for $x \leq 1$ samples at 300 K. Relatively high $T_N$ (~ 300 K) of $Bi_{1-x}La_xNiO_3$, compared with the other $ANiO_3$ series ($T_N$ = 225 K for $SmNiO_3$ is the highest), could be attributed to the divalent nature of Ni ions and/or the additional super-exchange interaction via Ni-O-$A$-O-Ni in which the role of ligand-hole is demonstrated explicitly.

The reentrant metallic behavior without anomaly in the thermopower implies that nanoscale islands of the insulating LT-phase grows in the matrix of the metallic HT-phase at low temperatures. Note that the insulating islands don't contribute to the thermopower. On decreasing temperature, growth of the islands freezes for the sake of gain of the elastic strain energy, accompanied with a large volume change, just like in relaxor ferroelectrics.[30] We call this coexisting state (insulating icebergs on the metallic sea) 'ligand-hole inhomogeneity'. These insulating icebergs are too small to be detected by X-ray diffraction. In the composition with $x$ = 0.05, the islands grow into the continent surrounding the metallic sea, making the system insulator and the triclinic perovskite. The large thermal hysteresis in resistivity could be attributed not only to the first-order character of the phase transition but also to the local fluctuation induced by martensitic-like character of transformation between LT- and HT-phase. The similar hysteretic behavior has also been reported for $PrNiO_3$ and $NdNiO_3$ (Ref. 11,12,31).

The aforementioned results are summarized in a schematic phase diagram shown in Figure 8. The phase boundary between the triclinic (insulating) and the orthorhombic (metallic) phase was extrapolated from the structural transitions of $BiNiO_3$ at ($P$, $T$) = (0, 513) and (3, 300), leading to a rate $\partial T_{MI}/\partial P$ = -71 K. In the case for $ANiO_3$ ($A$ = Pr, Nd), the rate $\partial T_{MI}/\partial P$ = -42 K and $\partial T_{MI}/\partial P$ = -76 K have reported by Obradors *et al*. and Canfield *et al*., respectively.[11,12] Given that melting of CD in all series of $ANiO_3$ ($A$ = Bi, Pr, Nd, …) is a phase transition from a solid to a liquid state of ligand-holes in 2$p$ orbitals, the pressure effect on $T_{MI}$ can be qualitatively explained with the Clausius-Clapeyron equation written as $(\partial P/\partial T_{MI})_{\Delta G} = \Delta S/\Delta V$. On the phase transition from a solid state to a liquid state of the ligand-holes, $\Delta S$ > 0 (an entropy change from an ordered state to a disordered state is positive) and $\Delta V/V$ = -2.5 % for BiNiO$_3$ and -0.25 % for PrNiO$_3$ (Ref.32) are given, leading to the inequality, $(\partial T_{MI}/\partial P)_{\Delta G} < 0$, which is consistent with the behavior that $T_{MI}$ decreases with increasing pressure. From the thermodynamic point of view, melting of the CD in $A$NiO$_3$ ($A$ = Bi, Pr, Nd, …) is phenomenologically similar to a phase transition from the ice to the water near the triple point.

In summary, we found that pressure ($P \geq 3$ GPa) and La-substitution ($x \geq 7.5$ %) suppressed the A-site CD of BiNiO$_3$ in essentially the same manner at room temperature, which was detected by the structural transformation from the triclinic to the orthorhombic perovskite accompanied with a steep drop of resistivity. In addition, temperature-induced melting of the CD in Bi$_{0.95}$La$_{0.05}$NiO$_3$ has clearly observed near 340 K. From the detailed structural studies combined with the transport and the magnetic measurements, we describe the charge distribution of the orthorhombic phase as [$A^{3+}\underline{L}^{\delta}$, Ni$^{2+}\underline{L}^{1-\delta}$]. Finally, a $P$-$T$ phase diagram of BiNiO$_3$ indicating a critical point (~ 3 GPa and 300 K) is given, which implies a general picture of ligand hole dynamics in $A$NiO$_3$. A large inhomogeneous region exists adjacent to both the metallic and the insulating regions. The present material provides a new intriguing aspect of perovskite oxides in the sense that strong covalency of Bi ion at A-site causes the MI transition dominated by fluctuation of the ligand-hole between A-site and B-site.

## ACKNOWLEDGMENT

The authors thank M. Lee, T. Mizokawa, and H. Wadati for fruitful discussions. This work is supported by the MEXT of Japan for Grants-in-Aid for Scientific Research A14204070, Grants-in-Aid for COE Research on Elements Science, Grants-in-Aid for 21$^{st}$ Century COE Programs at Kyoto Alliance for Chemistry. The synchrotron radiation experiments were performed at the SPring-8 with the approval of the Japan Synchrotron Radiation Research Institute.

**Figure captions**

Fig.1 XRD patterns of BiNiO$_3$ at 0.5, 1.1, 2.1, 3.0, 3.7 and 4.0 GPa with $\lambda$ = 0.4966 Å (a), Bi$_{0.95}$La$_{0.05}$NiO$_3$ at 100, 200, 300, 320, 330, 340, 360, 380 and 400 K with $\lambda$ = 0.776 Å (b), Bi$_{1-x}$La$_x$NiO$_3$ ($x$ = 0, 0.05, 0.075, 0.1, 0.2 and 0.5) with Cu K$\alpha$ at room temperature (c), traversing phase transition between triclinic (with CD) and orthorhombic (without CD) perovskite. An impurity phase was indicated by an asterisk.

Fig.2 Measured (+), calculated (line) and differential (bottom line) SXRD patterns for Bi$_{0.95}$La$_{0.05}$NiO$_3$ at (a) 400 K and (b) 100 K. The ticks indicate the positions of the reflections: upper in (a), orthorhombic; lower in (a) and (b), triclinic. The mass fraction of a triclinic phase at 400 K is 6 %.

Fig.3 Pressure dependence of (a) lattice parameters, (b) unit cell angles, (c) unit cell volume ($Z$ = 4), and (d) resistivity of BiNiO$_3$ at room temperature ($T$ = 300 K). In (a)-(c), the data were collected with increasing pressure. open and closed symbols signify the AP- and the HP-phases, respectively. Dashed lines in (b) are guides to the eyes.

Fig.4 Temperature dependence of (a) lattice parameters, (b) unit cell angles, (c) unit cell volume ($Z$ = 4), and resistivety of Bi$_{0.95}$La$_{0.05}$NiO$_3$. In (a)-(c), the data were collected on heating. Open and filled symbols signify the LT- and HT-phases, respectively.

Fig.5 La-content, $x$, dependence of (a) lattice parameters, (b) unit cell angles, (c) unit cell volume ($Z = 4$), and resistivety of $Bi_{1-x}La_xNiO_3$ at room temperature ($T = 300$ K). Dashed lines in (b) are guides to the eyes.

Fig.6 (Color online) Temperature dependence of (a) resistivity, (b) Seebeck coefficient, (c) magnetic susceptibility, and (d) inverse magnetic susceptibility of $Bi_{1-x}La_xNiO_3$.

Fig.7 Magnitude of the orthorhombic distortion, $b/a$, of $A$NiO$_3$ as a function of the mean ionic radii of A-site ion $r_A$ (a), and that of $Bi_{1-x}La_xNiO_3$ as a functions of La-content $x$ and pressure (b). The ionic radii for trivalent ions with 8 coordination were adopted for $r_A$.

Fig.8 Schematic phase diagram of $Bi_{1-x}La_xNiO_3$ as functions of pressure, La-content $x$, and temperature.

TABLE I. Structural parameters together with reliability factors for $Bi_{1-x}La_xNiO_3$.

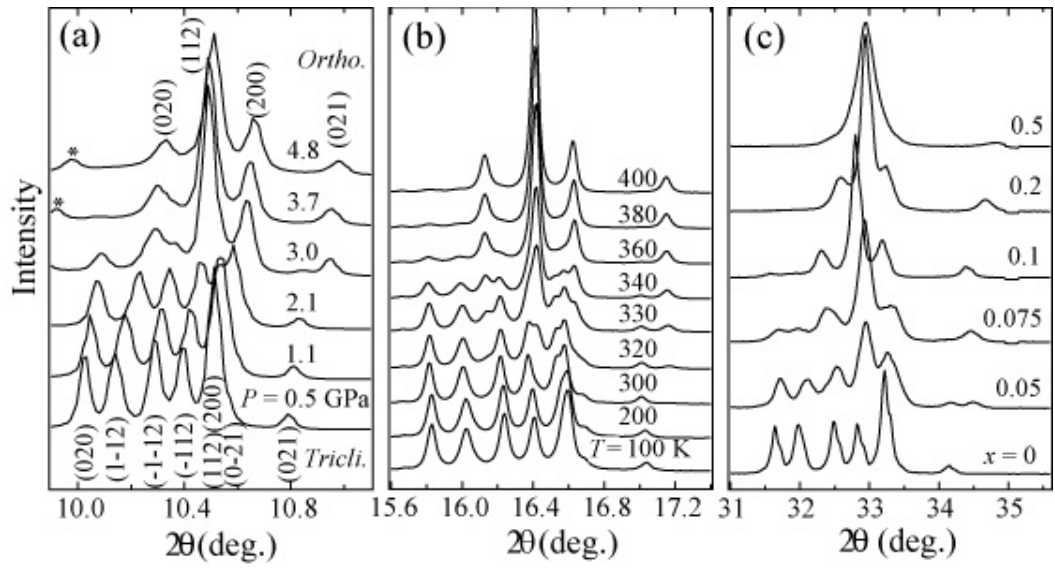

FIG. 1

S. Ishiwata et al.

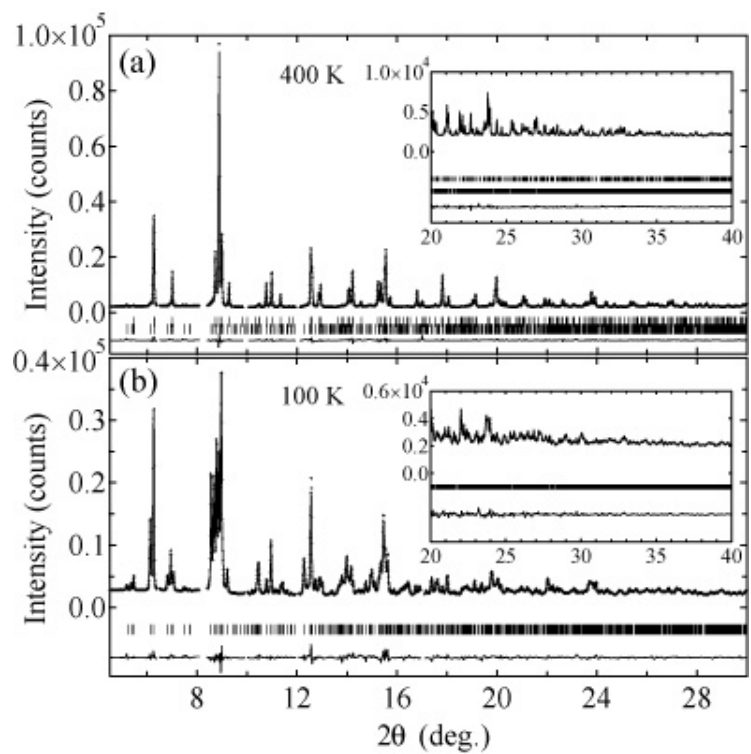

FIG. 2

S. Ishiwata et al.

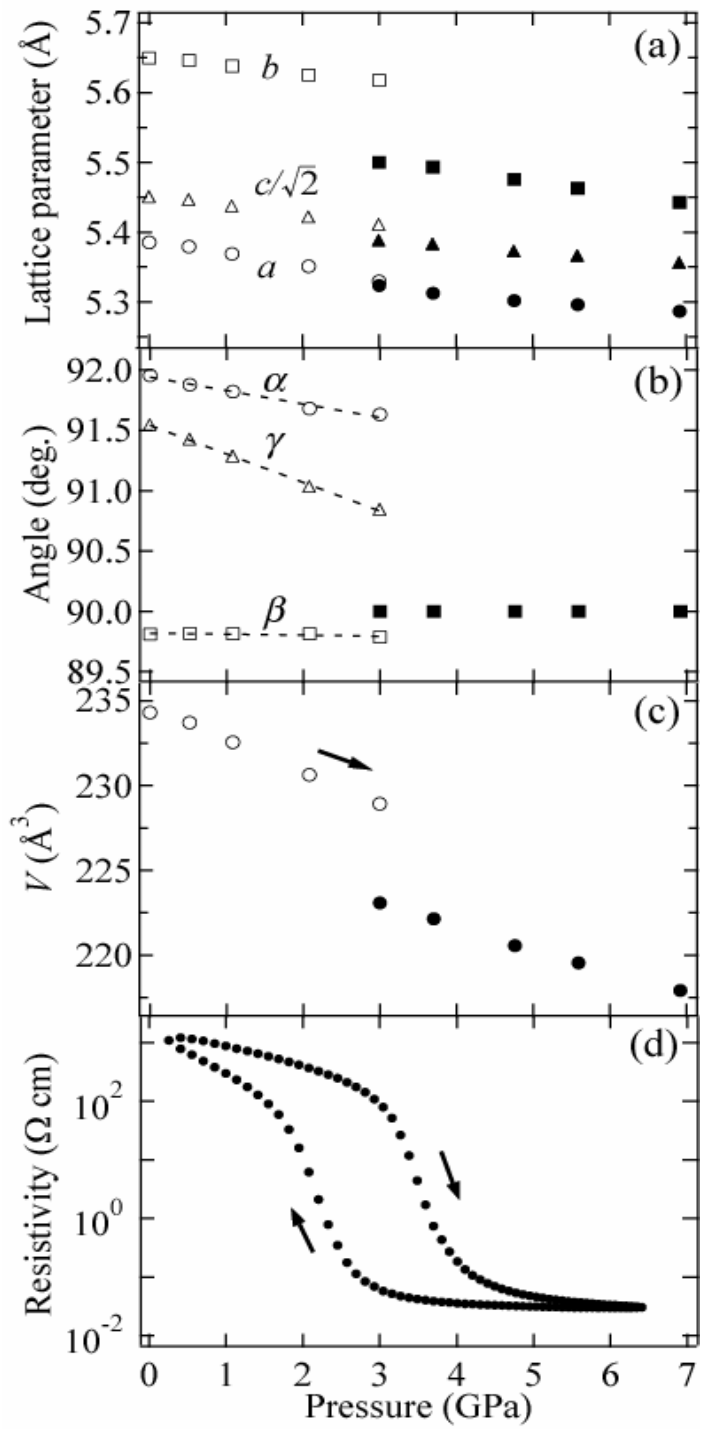

FIG. 3

S. Ishiwata et al.

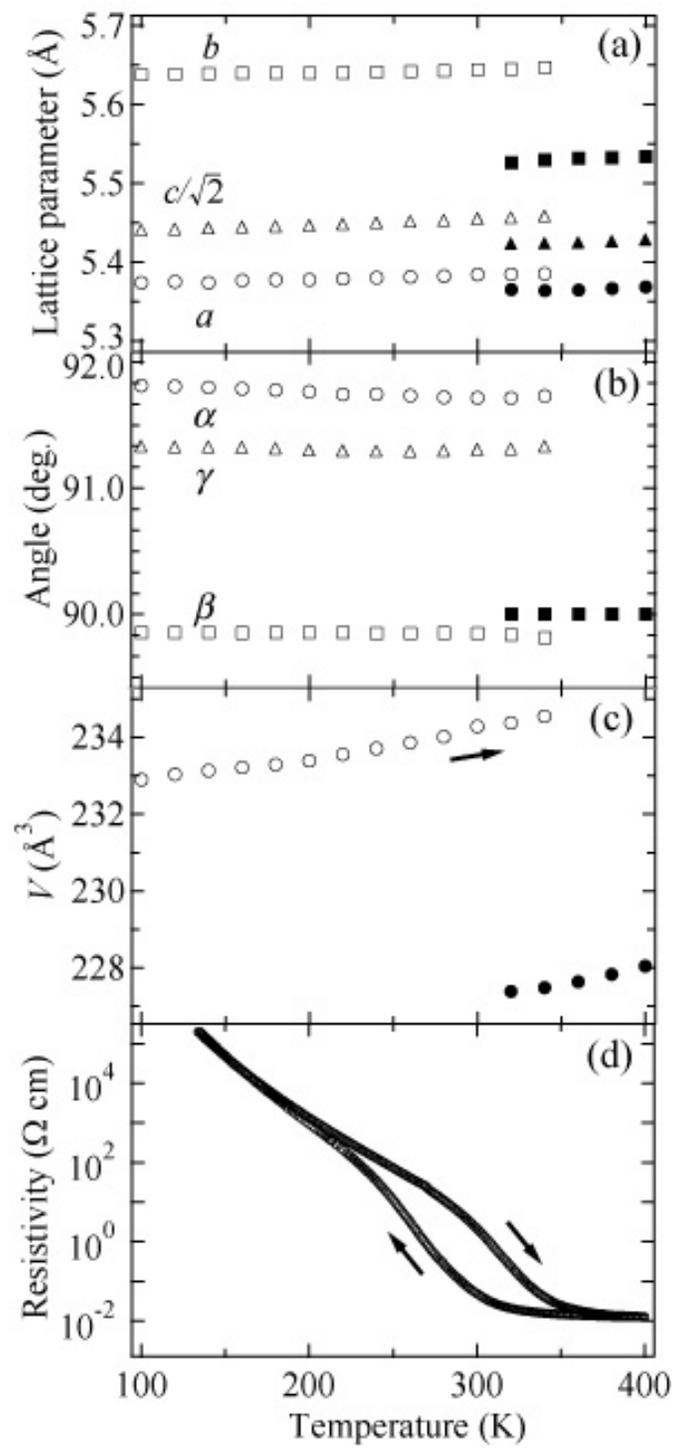

FIG. 4

S. Ishiwata et al.

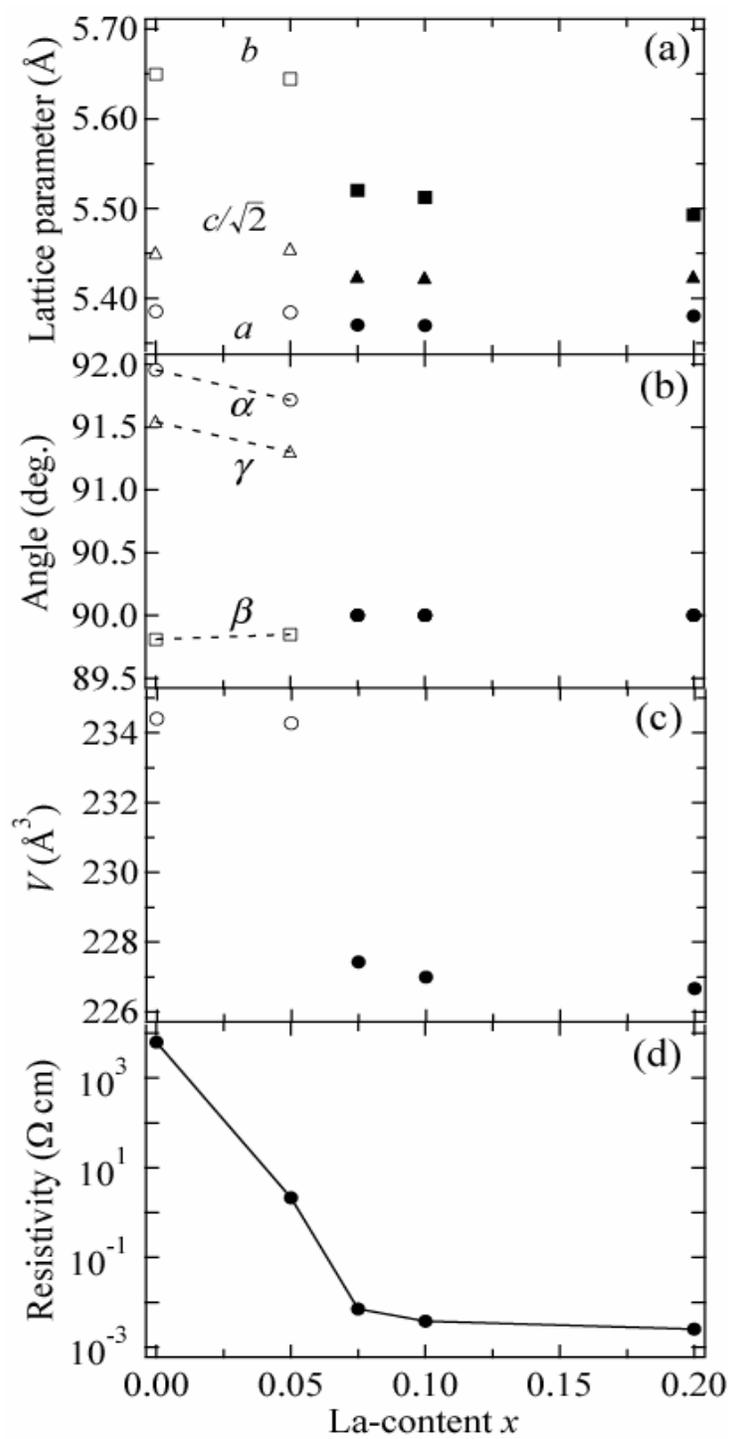

FIG. 5

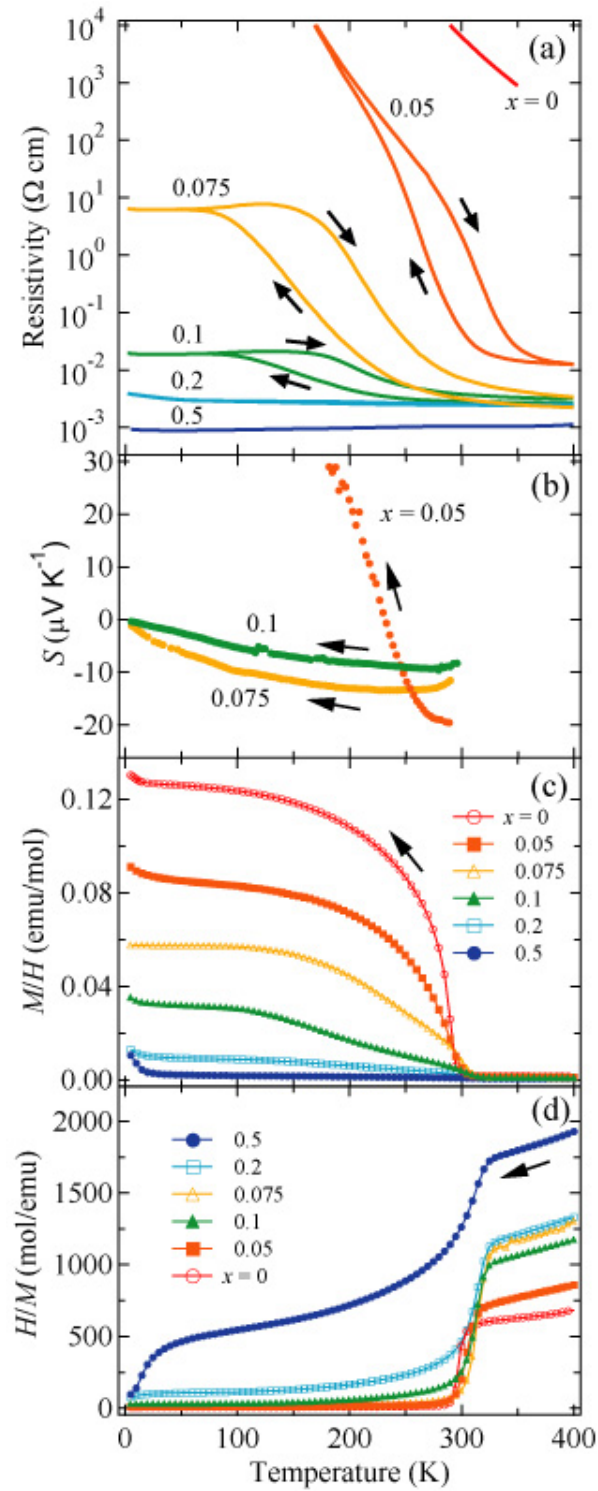

FIG. 6　　　　　　　　S. Ishiwata et al.

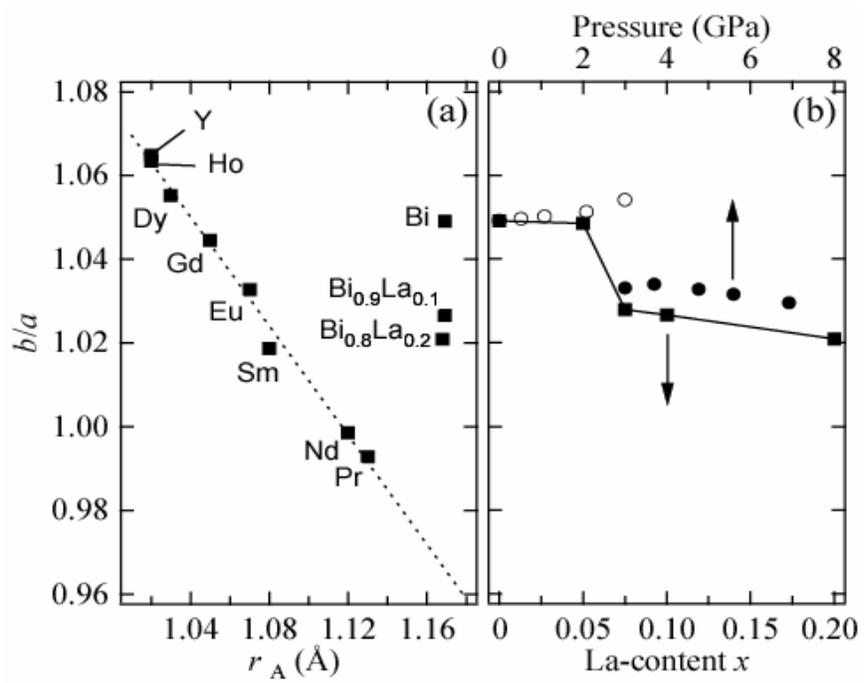

FIG. 7

S. Ishiwata et al.

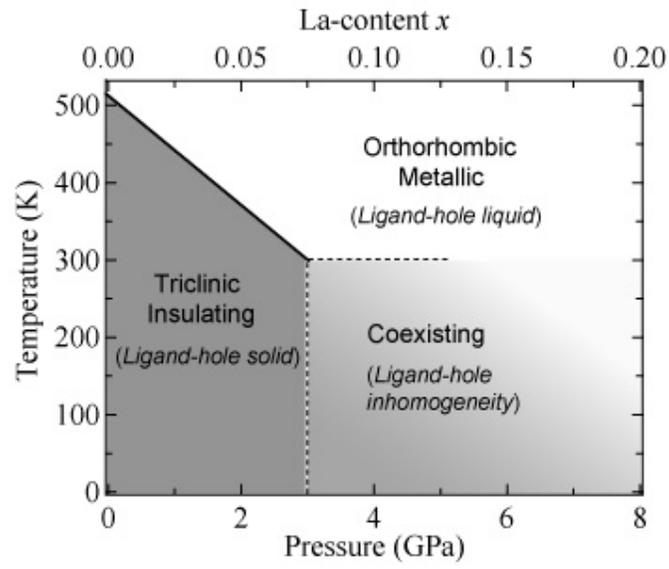

FIG. 8

| Parameters (unit) | (i)[a] | (ii) | (iii) | (iv) | (v) |
|---|---|---|---|---|---|
| | $x = 0$ | $x = 0.05$ | $x = 0.05$ | $x = 0.1$ | $x = 0.2$ |
| $T$ (K) | 300 | 100 | 400 | 300 | 300 |
| $a$ (Å) | 5.3852(2) | 5.3732(2) | 5.36723(6) | 5.36961(7) | 5.38025(10) |
| $b$ (Å) | 5.6498(2) | 5.6372(2) | 5.53200(6) | 5.51255(7) | 5.49285(9) |
| $c$ (Å) | 7.7078(3) | 7.6941(3) | 7.67412(9) | 7.66868(11) | 7.67017(16) |
| $a$ (deg) | 91.9529(10) | 91.792(2) | - | - | - |
| $b$ (deg) | 89.8097(9) | 89.853(2) | - | - | - |
| $g$ (deg) | 91.5411(9) | 91.307(2) | - | - | - |
| $V$ (Å$^3$) | 234.29(1) | 232.88(1) | 227.437(5) | 226.995(5) | 226.676(8) |
| $<A^{3+}-O>_8$ (Å) | 2.564 | 2.541 | 2.492 | 2.495 | 2.500 |
| $<A^{5+}-O>_6$ (Å) | 2.135 | 2.165 | - | - | - |
| $<Ni-O>_6$ (Å) | 2.091 | 2.078 | 1.986 | 1.980 | 1.973 |
| $<Ni-O-Ni>$ (deg) | 137 | 138 | 151.4 | 152.2 | 153.4 |
| $R_{wp}$ (%) | 2.72 | 3.89 | 3.39 | 3.71 | 3.90 |
| $R_I$ (%) | 1.35 | 0.91 | 1.90 | 0.82 | 1.19 |
| Space Group | $P$ -1 | $P$ -1 | $P\ bnm$ | $P\ bnm$ | $P\ bnm$ |

TABLE I

---

[a] From Ref. 18.

# References

[*] Present address: Department of Applied Physics, Waseda Univ., Ookubo, Shinjuku, Tokyo 169-8555, Japan. E-mail address: ishiwata@htsc.sci.waseda.ac.jp

[1] C. H. Chen, S-W. Cheong, and H. Y. Hwang, J. Appl. Phys. **81**, 4326 (1997).

[2] Y. Tokura and Y. Tomioka, J. Mag. Mag. Matter. **200**, 1 (1999).

[3] N. Ichikawa, S. Uchida, J. M. Tranquada, T. Niemöller, P. M. Gehring, S.-H. Lee, and J. R. Schneider, Phys. Rev. Lett. **85**, 1738 (2000).

[4] T. Yamauchi, Y. Ueda, and N. Mori, Phys. Rev. Lett. **89**, 057002 (2002).

[5] M. Takano, N. Nakanishi, Y. Takeda, S. Naka, and T. Takeda, Mater. Res. Bull. **12**, 923 (1977).

[6] P. M. Woodward, D. E. Cox, E. Moshopoulou, A. W. Sleight, and S. Morimoto, Phys. Rev. B **62**, 844 (2000).

[7] J. A. Alonso, J. L. García-Muñoz, M. T. Fernández-Díaz, M. A. G. Aranda, M. J. Martínez-Lope, and M. T. Casais, Phys. Rev. Lett. **82**, 3871 (1999).

[8] J. A. Alonso, M. J. Martínez-Lope, M. T. Casais, J. L. García-Muñoz, M. T. Fernández-Díaz, M. A. G. Aranda, Phys. Rev. B **64**, 094102 (2001).

[9] T. Saito, M. Azuma, E. Nishibori, M. Takata, M. Sakata, N. Nakayama, T. Arima, T. Kimura, and M. Takano, Physica B **329-333**, 866 (2003).

[10] M. Medarde, J. Phys.:Condens. Matter **9**, 1679 (1997).

[11] X. Obradors, L. M. Paulius, M. B. Maple, J. B. Torrance, A. I. Nazzal, J. Fontcuberta, and X. Granados, Phys. Rev. B **47**, R12353 (1993).

[12] P. C. Canfield, J. D. Thompson, S-W.Cheong, and L. W. Rupp, Phys. Rev. B **47**, R12357 (1993).